# Aesthetical criterion in art and science

Miloš Milovanović

Mathematical Institute of the Serbian Academy of Sciences and Arts, Belgrade, Serbia

milosm@mi.sanu.ac.rs

Gordana Medić-Simić

University of Belgrade, Belgrade, Serbia

gmedicsimic@gmail.com

**Abstract**

In the paper, the authors elaborate some recently published research concerning the originality of artworks in terms of self-organization in the complex systems physics. It has been demonstrated that the originality issue such conceived leads to the criterion of a substantial aesthetics whose applicability is not restricted to the fine arts domain only covering also physics, biology, cosmology and other fields construed in the complex systems terms. Moreover, it is about a truth criterion related to the traditional personality conception revealing the ontological context transcendent to the gnoseological dualism of subjective and objective reality that is characteristic of modern science and humanities. Thus, it is considered to be an aesthetical criterion substantiating art and science as well as the other developments of the postmodern era. Its impact to psychology, education, ecology, culture and other humanities is briefly indicated.

Since its beginnings aesthetics has been identified with the philosophy of art which implies that it has not only to speak about art and to produce judgments about artwork but has also to give a definition of what art actually is. There are two basic orientations in considering the issue – art as action and art as knowledge notwithstanding that aesthetics is neither ethics nor epistemology (Gethmann-Siefert, 1995, Croce, 1921, Weitz, 1956). Consequently, art could be specified as in-action or in-knowledge which does not imply that it is not an action or knowledge, but only that it is not subjected to definitions of ethics and epistemology requiring an expanded ontological context. Eventually, in-action also signifies an action, as well as in-knowledge is also knowledge – paraphrasing Socrates. In the expanded context, the very separation into ethics and epistemology seems to be a matter of concern.

The term stems from *aesthesis* (in Greek αἴσθησις) meaning a sensation, but also implying a personal experience. It was appropriated the first time in the current sense of a philosophical discipline by Alexander Gottlieb Baumgarten (1750) who was the follower of Gottfried Wilhelm Leibnitz (1720) in exploring correspondence between sensual and conceptual reality. Leibnitz' concept of perfection was also succeeded by Christian Wolff (1755) who realized it as truth, indicating an ontological context that becomes particularly relevant in considering the originality issue.

In the current paper, the authors consider the issue justifying some recently published research concerning the originality of artworks in terms of self-organization in the complex systems physics. The basic definition of self-organization originates from Shalizi (2004) using the concept of statistical complexity introduced by Grassberger (1986). Milovanović et al. (2013) have developed an effective algorithm for quantifying self-organization in large signals based on decomposing the signal using wavelet transform. Applied to artworks, it has been demonstrated that self-organization regularly occurs corresponding to the creative process which results in ability to identify forgery in art through the complexity comparison (Rajković, 2015). The paper elaborates the aforementioned results in order to elucidate their ontological context.

The exposed conception of art, however, transcends the viewpoint of modern humanities based upon an individualistic uptake of originality (Feyerabend, 1984). It reveals aesthetics figured through inspiration whereat the originality is established in terms of self-organization, implying a process that is not of individual character. By such a paradigmatic art the authors consider medieval ecclesial art referring exactly to the Orthodox iconography of Byzantine style.[1] A recent study has confirmed that in the iconography self-organized emergence is manifested through the form of personalized geometry signifying its intrinsic focus (Milovanović, 2015). The traditional concept of personality whose integrity is established through communion (Lossky, 2003, p.136) is actually based upon the conception of originality. On the other hand, since corresponded to self-organization, such aesthetics is not the only philosophy of art but also the philosophy of science. A paradigmatic science in that regard is considered to be just the complex systems physics.

The computational neuroscience related to the complex systems physics (Siegelmann, 2010) that way conjugates the access with neuroaesthetics (Pearce, 2016) indicating an opportunity of its foundation in an accurate manner. Such a computation should correspond to the neural substrates of aesthetics relating sensation and neural activity that is the crucial aim of psychophysics in Fechner`s (1860) view referred to as *inner psychophysics* (Boring, 1950; Scheerer, 1987). Its elaboration warrants an expanded ontology that is in line with both art and science, realizing aesthetics not restricted to the fine arts domain only, but intrinsic to some objects, activities, or occasions people hold most dear to them as a vital component that contributes to heighten their uniqueness and specialness (Dissanayake, 1988; 2009). A transcendental experience shaped following strong emotional arousal states relates it to religious beliefs implying embodied cognition as one`s procedural memory (Soliman, 2015). Cultural particularities commonly construed in terms of different biases for information processing (Park, 2010) are interwoven through the ontological context concerning a religious tradition embodiment. The Tabor light experience pervasive to the Orthodox iconography is regarded to be a testimony of the aforesaid.

---

[1] Byzantine style is a technical term not corresponding to any artistic style. It refers to the Orthodox ecclesial art that has occurred in its integrity after iconoclastic period in the Eastern Roman Empire finding its dogmatic foundation by the Seventh Ecumenical Council in Nicea.

The matter is exposed in the following way. In Section 1 the subject and method are presented, as well as the order of exposing the matter. Section 2 concerns creativity in art relating it to the intrinsic time conception. In Section 3, the basic conceptions of complexity and causality in their interrelationship with aesthetics are given. Section 4 considers the originality in terms of self-organization in artworks. In Sections 5 and 6 there are presented two paradigms of such aesthetics – artistic and scientific one. Section 7 contains some concluding remarks.

**Time and creativity**

The issue of time in artworks was fundamentally discussed in a paper published by Etienne Souriau (1949). The article principally refers to the plastic arts (design, painting, sculpture, architecture) but the author suggests that the artworks in general could be treated in the same way.

At the very beginning of the article, it is claimed that…

…nothing is more dangerous for exact and delicate understanding of plastic arts than their rather banal description "arts of space" in contrast to phonetic and cinematic arts characterized as "arts of time".

This contrast, subscribed to by a great number of aestheticians from Hegel to Max Dessoir, has its historic origin in the philosophy of Kant, particularly in the contrast he makes between the external senses, to which the form of space would be inherent, and the internal sense whose form would be time. The desire to bring music and poetry into the realm of the internal sense (in order to see there "the soul speaking directly to the soul") has often led to a real misunderstanding of the extent and the cosmic reach of the plastic arts, stripped of their temporal dimensions, and of their content according to that dimension.

The widely accepted distinction also popularized by Lessing is considered to be flawed as it is unable to account the complex nature of arts (Antonova, 2010, p.5). To overcome the evident deficiency of the approach Souriau introduces the concept of the *intrinsic time* defining it as 'artistic time inherent in the texture itself of a picture in its composition, in its aesthetic arrangement' (Souriau, 1949, p.296-297). His view could be regarded as a starting point for the conception being elaborated in the current paper.

The postulated conception contemplates art as a creative process which applies both to the production and to the observation of an artwork. It implies an inherent organizing principle of the artwork actually corresponding to self-organization in terms of the complex systems physics. Rudolf Arheim (1971, p.29-30) qualified creative process as self-regulatory in the essay published in 1971, before the concept of self-organization was recognized in exploring of far-from-equilibrium dynamical systems (Nicolis, 1977). In the theory developed somewhat later by Ilya Prigogine (1980), self-organization in dynamical system is related to the existence of the intrinsic time operator, which is exactly corresponded to Souriau`s conception of time mentioned above.

The creativity, such perceived becomes synonymous with originality impersonating a processual designation of art. The paper by Rajković et al. (2015) concerns creativity in art giving a method for recognition of creative process based upon quantifying self-organization. The method considers an artwork as an element of the signal space decomposing it in a hierarchical basis generated through translation and dilatation of a single special function termed wavelet (Mallat, 2009). The representation in such a basis enables establishing a statistical model of the signal, termed hidden Markov model (Crouse, 1998) with correlations recognized through the form of a causal structure. The global complexity defined as Shannon information contained in the causal structure corresponds to quantifying self-organization (Milovanović, 2013). In that manner, the representations in various bases become subjected to the complexity comparison considering the optimal one is the basis in which the global complexity is maximized. It is generated by the optimal wavelet identifying in the best way the inherent organization of the signal. Furthermore, the complexity of a signal is regarded as maximal complexity over all hierarchical bases each of them corresponded to an intrinsic time operator on the signal space (Milovanović, 2015a). The art conceived in that way is realized through encryption and decryption process (Svozil, 2008) implying an optimal communication specified by maximization of complexity.

The method exposed is corroborated by some experimental results regarding the originality issue. The title of the paper 'Artists who forged themselves' is already indicating that it is about replication of an artwork by the author himself. The data set consisting of seven high resolution images painted by the Dutch artist Charlotte Caspers is used. The artist was commissioned by Ingrid Daubechies and the members of the Machine Learning and Image Processing for Art Investigation Research Group at Princeton University to paint 7 paintings of relatively small size (approximately 25 cm x 20 cm) of different styles and using different materials. Within the next few days she has also painted a replica of each painting using the same paints, brushes and grounds and under the same lighting conditions (Fig.1). For the method and the results presented, it is of interest to mention the remark of Daubechies that Caspers spent close to two times more time on creating each replica as compared to the original, indicating that 'painting a copy is a more painstaking process than the spontaneous painting of an original'. The copies were of such high quality that the artist was convinced that one would not be able to distinguish them from the originals (Daubechies, 1985). However, it has been demonstrated that all the original paintings in the optimal wavelet decomposition have higher global complexity than replicas of the same (Rajković, 2015). In such a way, the forgery in art could be detected through complexity comparison of the artworks having known that one of them is a replica of another.

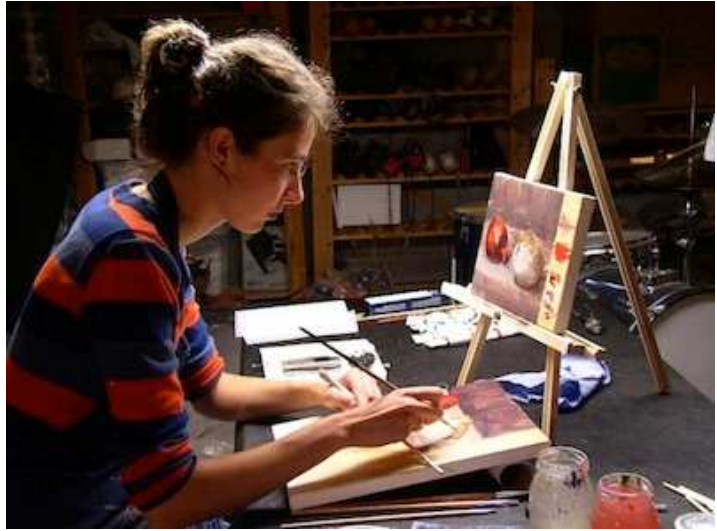

Figure 1. Dutch artist Charlotte Caspers replicating her own artwork

According to the definition of global complexity, it could be concluded that the replication process which reflects stronger causal necessity than the spontaneous, original creation corresponds to less informative causal structure determining the complexity of the model. In other words, the complexity of the model represents what is being appointed as an overcome of causal necessity. Related to that, Zausner (2007) has written: 'Creating a work of art is an irreversible process involving increasing levels of complexity and unpredictable events.' The increase of complexity in temporal domain is actually the definition of self-organization originating by Shalizi (2004).

 **Complexity and causality**

In 1990, Jürgen Schmidhuber postulated an algorithmic theory of aesthetics. Using the deterministic computation theory, he related aesthetical pleasure to minimal description length stating that among several observations classified as comparable by a given subjective observer, the aesthetically most pleasing one is with the shortest description given the observers previous knowledge and his particular method for encoding the data. A paradigmatic form of such aesthetics could be less detailed XV century proportion studies by Leonardo da Vinci and Albrecht Durer describing an aesthetically pleased human face with very little information. The premise is that any observer continually tries to improve the predictability and compressibility of the observations by discovering regularities such as symmetry and self-similarity (Schmidhuber, 1997).

The Schmidhuber`s approach is actually the definition of algorithmic complexity relating the aesthetic optimality to its minimization. In that regard, it corresponds to the conception of universal Turing machine (Crutchfield, 1989a) that is a general model of deterministic computation and causal necessity. However, the main problem with algorithmic information theory is that it deals with possible, but not with probable, although some practitioners also apply ad hoc some probabilistic concepts (Grassberger, 1986, p.908). For aesthetical considerations, this is disastrous if having in mind that the creativity in art corresponds to an overcome of causal necessity. Therefore, instead of algorithmic information theory, statistical information theory should be required as well as statistical causality instead of deterministic one. Through such an approach the algorithmic complexity defined in terms of universal

Turing machine is supposed to be replaced by statistical complexity in terms of Bernoulli-Turing machine that is related both to deterministic and to stochastic computation (Crutchfield, 1989a).

The concept of statistical complexity is introduced the first time by Peter Grassberger (1986). Modelled by the human ability to make abstractions, he founded the measure of complexity corresponding to the Shannon information needed to describe properties of ensembles of patterns. For instance, when shown pictures of animals, one immediately recognizes the concepts 'dog,' 'cat,' etc., although the individual pictures showing dogs might in other respects be very different. So one immediately classifies the pictures into sets, with pictures within one set considered to be equivalent. Thus, one actually has ensembles – when calling a random pattern complex or not, one actually means that the ensemble of all similar patterns is complex or not complex. Grassberger`s conception is subsequently developed by Crutchfield and Young but the basic idea already contains an aesthetic moment related to human sensation.

Although Grassberger identified the complexity of a system with the information needed to specify the state of its optimal predictor, he still did not give any indication about of how such a predictor and its state might be found nor even what the optimal prediction might mean (Shalizi, 2001). Nevertheless, he discerned that a common characteristic to all instances of complex systems mentioned in the paper is that the complexity is self-generated in the sense that the formulation of the problem is translationally invariant and the observed structure arises from a spontaneous breakdown of the symmetry. Crutchfield and Young (1989) extended his conception by giving accurate definitions of optimal predictor and state. The basis of their approach is an abstract concept of complexity signifying the information contained in minimum number of equivalence classes obtained by reducing a data set modulo a symmetry. In that manner the concept of causal state has been established corresponding to the intrinsic computability of a process in terms of Bernoulli-Turing machine (Chrutchfield, 1989a) that was used by Shalizi (2004) to define self-organization in an accurate manner.

According to his view, self-organization signifies the increase of statistical complexity in temporal domain which is also reflected by the causal structure of a process. The causal structure actually contains minimal information required for optimal prediction both globally and locally, thus being factorized through spatio-temporal lattice onto local causal states in a form of Markov random field (Shalizi, 2001). The increase of local complexity implies that causality is becoming more informative, indicating an overcome of causal necessity since Shannon information is the entropic measure corresponded to the uncertainty of a random variable. In other words, one needs more and more information to specify the causal state of a system in opposition to the causality figured through necessity containing no information. The global complexity signifying information contained in the causal structure is supposed to be an indicator of self-organization (Milovanović, 2013).

**Self-organization and originality**

Although the conception of spontaneous organization is very old it crystallized into the term self-organization in circles connected with cybernetics and computer science (Shalizi, 2001,

p. 6). The first appearance of the term seems to be in 1947 by William Ross Ashby, who was a British psychiatrist active in the cybernetics movement after the Second World War. He understood a system to be self-organizing if it changed its own organization, rather than being rewired by an external agency. Ashby`s ingenious answer to how it is possible has been that it is not. As governed by deterministic computation theory, he was holding the view that organization should be an invariant of the system stated by a functional dependence. Fortunately for the conception, Ashby`s idea of what constitutes self-organization have been pretty thoroughly ignored by everyone else who`s used that (Shalizi, 2001, p. 7).

A significant shift in its realization took place due to the Brussel School of Thermodynamics headed by investigations of Ilya Prigogine. The intrinsic time operator formalism (Prigogine, 1980) in their access appears to be an effective criterion of what the complex systems physics should be. A correspondence between self-organization and the intrinsic time appears to be a fundamental perception of statistical causality in complex systems. The definition of self-organization by Shalizi corresponded to the increase of statistical complexity in the temporal domain perfectly fits into the context concerning that it implies the time conception not demanded about its origin. The access, however, clearly states the significance of time axis through overcome of causal necessity, assuming by that an informative causal structure corresponded to the global complexity of the system.

In that manner, self-organization transcends the conception of causality corresponded to modern science disclosing the ontological context that appears as indicative for aesthetical considerations. Moreover, such aesthetics does not fit into the framework of ethics and epistemology whose positive definitions are inflicted by the causality conception. Consequently, one should specify it in terms of in-action or in-knowledge implying by that an expanded ontological context not subjected to the modern humanities. One discerns concurrence with Jung`s analytical psychology wherein the emergence of significant coincidences without an apparent cause is regarded as an expression of a creative process due to the archetypal images substantiating time in terms of their action (Jung, 1989, p.107-109).

Regarding the originality issue, such aesthetics also transcend the individualistic uptake based upon modern humanities. Since established in terms of self-organization, it is figured through inspiration implying a process that is not of individual character. Conceived in that manner, the origin is implemented in its basic purport impersonating an archetypical content being manifested by the person concerned. This is actually what one takes under the term personality in the traditional sense. To be definite, the paper is confined to the tradition of the Orthodox Christianity and its uptake of personality whose integrity is established through communion (Lossky, 2003, p. 136).

Above all, one should be aware that the term comes from Latin word *persona* meaning a theatrical mask. In that content, it appears also in Jung`s analytical psychology signifying one`s social presentation. In the Orthodox Christianity, however, personality is never used in that sense. In order to designate a personality the Orthodox tradition uses the term *hypostasis* (in Greek ὑπόστασις) actually not to be subjected to the incorrect meaning. The Latin translation of the term is the word *substantia,* but it has never been used in the sense of the Orthodox tradition.

The traditional personality conception inevitably pertains the originality issue. This is emphatically reflected by Vladimir Lossky (2003, p. 49) in statement that personality is an opposite to the concept of individual.

Indeed, the conception of human personality as something substantial making of every human individual a unique being that is absolutely incomparable to anyone and irreducible to the other individual is exactly provided by the Orthodox tradition. The philosophy of the ancient world knew only the human individuals. Human personality cannot be conceptually reflected. It is shielded from any rational determination not allowing even a description since all the properties by which one makes an effort to characterize it are still present in other individuals. Personal substantiality in one`s life could be observed only through immediate intuition or presented through an artwork. Saying –This is Mozart, or – This is Rembrandt, one is positioned in a substantial area with no corresponding equivalent anywhere.

Lossky definitely states that a personality cannot be conceptualized. Furthermore, he emphasizes that it is consolidated only through the abdication of determining and enslaving by causal necessity (Lossky, 2003, p. 95). The abdication elucidates the character of in-action or in-knowledge designating the ontological context of originality. In order to recognize an authentic aesthetical criterion all the causal restrictions have to be overcome. Consequently, in the expanded context of self-organization positive definitions of ethics and epistemology lose their significance disclosing a requirement for their substantial designation in ontological terms based upon the traditional personality conception.

A famous Romanian conductor, composer and teacher Sergiu Celibidache professing his philosophy of art has remarked that the creative act accepts no conditions outside of itself. The overriding goal of art is to attain the freedom and that is why one says it is perfect. It is not beauty, but it is truth corresponding to the conception of the optimal basis for experiencing an artwork designated by him as the *central cosmic vibration.* Beauty is just a stage on the path to truth – to be able to perceive the sensation of the central cosmic vibration emerging from within oneself. The attained transcendental experience is what he has termed freedom in art (Celibidache, 2013).

What an incredible and invaluable opportunity is, over one and a half hours to attain freedom from the hand that terrorizes and belabors me.  That is what I want. I want to go beyond – That was beautiful, and say – That was good! I am free again.

Celibidache's approach to music-making is often described more by what he did not do instead of what he did (Oestreich, 2007). He was principal conductor of Munich philharmonic, Berlin Philharmonic and several European orchestras. Later in life, he taught at Mainz University in Germany and the Curtis Institute of Music in Philadelphia. But he frequently refused to release his performances on commercial recordings during his lifetime claiming that a listener in that way could not obtain a transcendental experience. Many of the recordings of his performances were released posthumously. In his insisting on the originality of an artistic performance, some recognize aspects of Zen Buddhism, such as *ichi-go ichi-e*, the term often translated as 'for this time only', 'never again' or 'one chance in a lifetime'. The term reminds one to cherish any gathering that he partakes in, citing the fact that it is unique and unrepeatable.

The statement his approach to music is not personal one is deduced considering that it demands an experience transcendent to all knowledge and any description. But precisely because of that it implies a substantial moment actually corresponding to the traditional conception of personality established through communion that is exactly a transcendental experience. In that manner, one could say that the crucial influence on his art and uptake of originality still belongs to the Orthodox Christianity that was actually his ancestry religion.

**Aesthetics of the Orthodox iconography**

The personal substantiality that is also referred to as the traditional personality conception constitutes the foundation of the Orthodox Christianity. The traditional gnoseology is actually based upon the substantial moment as well as aesthetics and ecclesial art. In that respect, the entire Orthodox iconography is perceived as a revelation of personality as it was formulated by resolutions of the Seventh Ecumenical Council in Nicea having ended the iconoclastic period in the Eastern Roman Empire. The ecclesial art occurred after the iconoclasm – technically designed as Byzantine style – has been manifested through a profound iconographic content which is the synthesis and the ultimate reach of the Orthodox tradition. Elucidating a substantial moment, the geometrical optics of icon are outlined by a dynamical spatial structure commonly referred to as the *reverse perspective*.

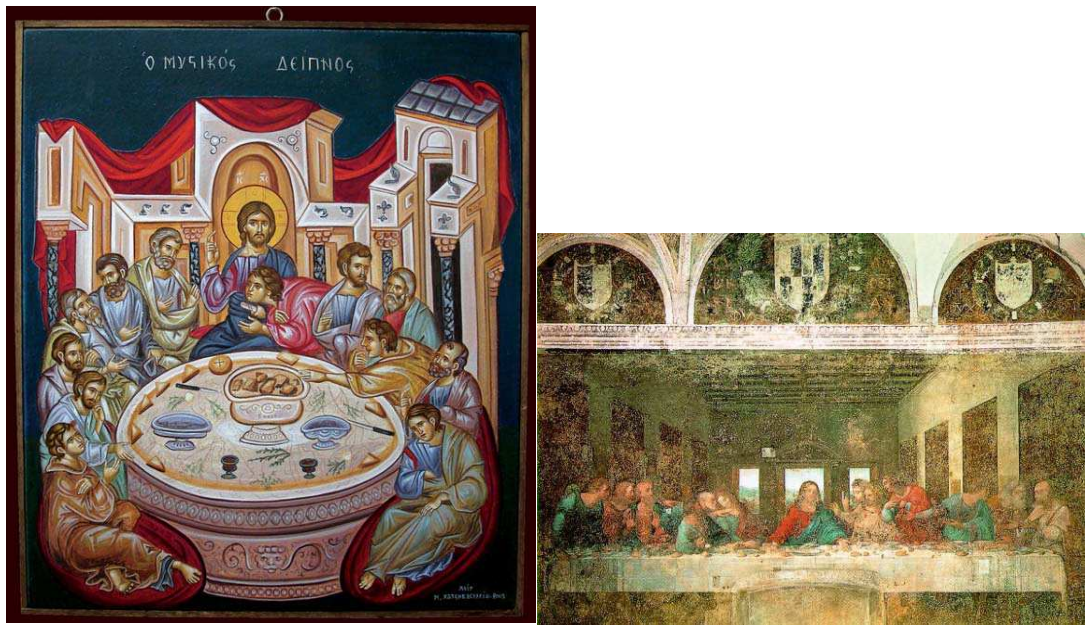

Figure 2 The last supper – left: Reverse perspective of the Orthodox icon representing expansion towards the interior – right: Linear perspective in the Renaissance image representing a contraction towards the interior

Although not clearly identified in all periods and artistic styles, reverse perspective is considered to be a defining feature of the Orthodox iconography (Antonova, 2010, p.29). Being like that it is not confined to painting only, although the special artistic form makes it particularly evident. It is actually about the image expansion towards the interior (Fig. 2 – left) that regularly occurs in traditional ecclesial art as a designation of a highlight quality disclosing the intrinsic time of an icon (Milovanović, 2015). The term is initiated in 1920s by Russian mathematician Pavel Florensky (1999) who established it in opposition to the modern linear perspective pointing out the image contraction towards the interior. It was the first mention of the concept which, notwithstanding it exists in iconography from olden times with fundamental significance, had never been theoretically discussed before, nor there had anything has been written about that even in iconographic canons. Almost twenty years later, the concept was also discussed by Serbian mathematician Miloš Radojčić (1940) under the term *counter perspective*. One should mention, however, that the both of terms are inappropriate since they are construed in respect to the linear perspective of modern spatiality that has become a dominant display mode only since the Renaissance (Fig. 2 – right).

Till nowadays there have been a few more or less successful interpretations of the concept. In reversing of linear perspective, some authors have realized removing material properties and the actual physical presence of the shapes (Lazarev, 2004, p.18) that has been denied afterwards (Džalto, 2008, p. 15). Some were discovering analogy with binocular vision in human perception that implies a complex imaging irreducible to the linear perspective mode (Antonova 2010, p. 33). The access could be summarized in the representation of a shape from different viewpoints in the same image that was a starting point for Clemena Antonova to ascertain the significance of time in iconography. Referring to Souriau (1949), she assumes time as a fundamental organizing principle of pictorial art (Antonova 2010, p.5) that indicates presence as a basic content of icon manifested through the attendance of various time instances in the same image qualifying different viewpoints. Although she identified the complex nature of icon, the deficiency of an adequate conceptual framework made her access incomplete in its final inference.

Complementing her considerations Milovanović et al. (2015) have developed a thorough access to iconography based on complex systems physics and fractal geometry concerning the intrinsic time conception recognized in the reverse perspective expansion of the image (Milovanović). The depth dimension established in that manner actually corresponds to the time axis in terms of self-organization signifying an overcome of causal necessity. In that respect, the personality dynamics, bringing a substantial moment to the fore is proven to be a fundamental organizing principle of the Orthodox iconography. This is what the authors refer to as *personalized geometry* of icon relating it to self-organization in the hidden Markov model of the artwork signal processing (Milovanović, 2013).

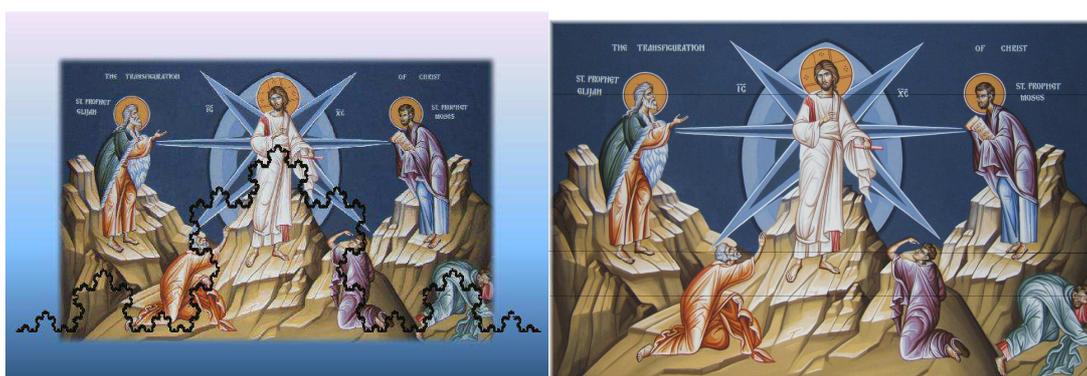

Figure 3 The Icon of Transfiguration – left: The emergence of the Von Koch curve disclosing a scaling property corresponded to the intrinsic time – right – The personalized geometry recognized through a progression from horizontal, through semi-vertical to finally vertical position in respect to the central Christ`s figure

The intrinsic time of such a model corresponds to dyadic scale in the wavelet decomposition of the signal space disclosing a scaling property that is discernible in the Icon of Transfiguration through the emergence of the Von-Koch curve (Figure 3 – left). Its existence in the image reveals the iconography as a paradigmatic art of the aesthetics designed at self-organization and statistical causality. Moreover, it is just a pattern of the personalized geometry arising through the reverse perspective expansion of mountain massifs towards the image interior. Thus establishing a spatial depth signifies a personality dynamics reflected in progression form the horizontal through semi-vertical to finally vertical position in respect to central Christ`s figure (Fig. 3 – right). In that manner, the geometry of the icon incident to its content represents communion in immaterial Tabor light corresponding to the personality liberation indicated by Celibidache`s (2013) philosophy of art. The Tabor light experience constitutes grounds for the mystico-ascetic practice of the Orthodox Christianity (Lossky, 2003, p.160) and so one considers equally the aesthetics of asceticism (Laziić, 2008) assuming the concept of liturgical time incident to the intrinsic time in iconography (Kalokyris, 1966). In that manner, the iconography is associated to the integral mythological content of the Orthodox tradition due to its aesthetical significance relating the ecclesial Synaxis.

Discussing time in mythology Claude Lévi-Strauss refers to de Saussure (1969, p. 117-119) and his consideration of language through two complementary aspects – *langue* and *parole* – of which the first one is subjected to the reversible and second one to the irreversible time conception. The myth is defined by a system that integrates both of them impersonating an original linguistic expression (Lévi-Strauss, 1955, p.430). Combining that way structural and statistical description in the designed system Lévi-Strauss (1955, p.443-444) actually states the complex nature of the myth that inevitably demands statistical causality in its discussion. The access exactly corresponds to the complex systems physics developed by Prigogine (1980) whose scientific engagement predominantly concerned the unification of reversible and irreversible time physical theories. The statistical causality appearing in both instances reveals their substantial interrelation figured through the traditional personality conception. As in considering the artworks the informational content of the causal structure also concerns the originality issue.

**Aesthetics of the complex systems physics**

The modern science has done its best to be separated from the mythological ancestry standing at the root of its emergence. In the final form, however, it is manifested as an utterly naïve mythological system apparently revealing its genealogy from the Western European scholasticism of the Middle Ages (Prigogine, 1982, p.47-49). The system appointed by Prigogine (1980, p. 214) as the founding myth of the classical science in its general shapes implies a deterministic vision inflicted by the causal necessity. It is entirely reflected through the modern spatiality having regard the linear perspective as its dominant display mode as well as through the heliocentric cosmology having placed the simplicity of the geometrical description to the fore (Milovanović, 2013b). Both instances of the system operate with no causal information since they are governed by a causality not allowing any complex behavior. In that manner could be taken Blake`s criticism of Newton and his mathematical physics that is actually the aesthetical one. In Newtonian mechanics William Blake has recognized a paradigm of a crude simplification that disclaims the complexity of nature and its aesthetics (Burwick, 1986, p.8) in favor of a geometrical idealization figured through Newton`s causal laws (Fig. 4).

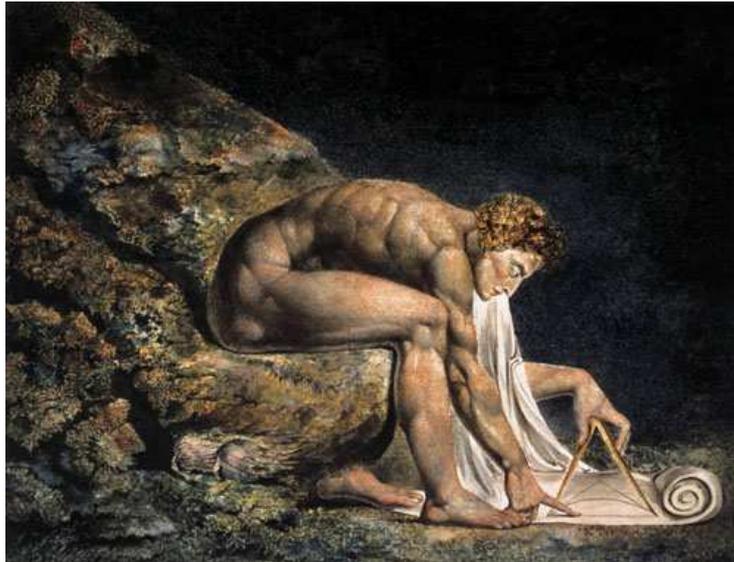

Figure 4. *Newton* – a monotype by William Blake. The painting represents his disclaiming the nature and its aesthetics in favor of a geometrical idealized

As early as 1754, d`Alambert noted in the *Encyclopédie* that one could regard the duration as a fourth dimension supplementing the common three-dimensional design (d'Alembert, 1754, p.1010) and Lagrange (1794) more than hundred years before Einstein and Minkowski (1909) went so far as to term it a four-dimensional geometry.[2] *Encyclopédie, ou dictionnaire raisonné des sciences, des arts et des métiers* edited by Denis Didrot and Jean-Baptiste le Rond d'Alembert was a general encyclopaedia published in France between 1751 and 1772, with later supplements, revised editions and translations. More than as collection of all the knowledge it served as a manifest of the New Age represented at the societal level by Industrial revolution. The introduction referenced as *Discours préliminaire* is considered to be the most important manifest of Enlightenment ideals (Milovanović, 2009). Reorganizing knowledge based upon human reason instead of by nature or theology the *Encyclopédie* opened a gate to the modern humanities incident to political secularism being occurred through the French revolution. But the most significant feature regarding to the aesthetical criterion concerns the time issue figured in terms of what one appoints to as its linear parametrization.

The claim that time is nothing but a linear parameter corresponds to Meyerson`s definition of the modern science history as a progressive realization of the fundamental bias in the human reasoning – reducing of difference and change to identity and constancy. The requirement appointed by him as *elimination of time* (Meyerson, 1930, p.215) is an immediate consequence of a causality conception destitute of any informational content having therefore no idea about self-organization. The climax of such a historical trail is represented by Albert Einstein and his categorical rejecting the existence of change in physics. The science in Einstein`s view was an attempt to go beyond the world of appearance to reach the timeless world of supreme rationality – the world of Spinoza (Prigogine, 1980, p.255). The conviction certainly related to his Jewish ancestry and religiosity (Naumann, 2015) was definitely manifested in celebrated Einstein-Bohr debate on the foundations of quantum theory. The core of the debate was exactly about the statistical causality conception and the fundamental role of probability in specifying a system state that was emphatically denied by Einstein who was supporting an objective epistemology of physics. Although he overturned Newtonian mechanics having unfolded new horizons of modern science, Einstein was firmly holding the Cartesian intent of reducing physics to geometry in terms of the modern spatiality that is nothing but the corresponding causality conception.

---

[2] The concept of duration such perceived corresponds to the psychological time conception also mentioned by Souriau (1949, p. 296) in opposition to the intrinsic one. Its basic representation concerning a metaphorical relationship to spatiality (Casasanto, 2008) is actually the content implied by the first Newton`s law that specifies a linear parametrization of time due to the inertial motion.

A recovery of time in physics had already been suggested by formulation of thermodynamics in XIX century. However, it has only quantum theory pointed out a statistical character of causal laws allowing the personality to be involved in a system through a process of observation and measurement. Ilya Prigogine (1980) has initiated a theory referred to as far-from-equilibrium dynamics that has placed the time issue to the fore. In order to give the time conception an original significance he has brought into the theory the intrinsic time operator corresponded to the observables in quantum theory acting on the state space. The operator inflicts a similarity transformation from deterministic Lie group symmetry to probabilistic Markov semigroup exposing statistical causality of the complex system physics. The equivalence of both descriptions of a system – deterministic and probabilistic one – revealed a probabilistic theory that is complete and objective wherewith the Einstein-Bohr debate should begin to take new shapes (Prigogine, 1980, p.255). The probability in the complex system physics becomes a fundamental feature of the causal structure concerning self-organization that is a substantial moment of the theory.

In that regard, one recognizes the complex systems physics as a mythology of science, realizing a substantial moment corresponded to the aesthetical criterion of the Orthodox iconography. According to that, it is considered to be a paradigmatic science of the same aesthetics. Furthermore, the common aesthetical criterion concerns the phrase: *I am the truth*, pronounced by Jesus Christ that is about the complex nature of the God-man revealing the divine personality.[3] It is also an argument about the originality of such a mythology in respect to the myth of the classical science. By comparing their complexities figured through the information contained in the respective causal structures one realizes that the classical science actually deals with no substantial moment that would constitute any causal information.

The aesthetical criterion conceived in that way consolidates both art and science overcoming the gnoseological dualism concerning subjective and objective reality that is also characteristic of the modern humanities. Moreover, it is actually the truth criterion constituted through the traditional personality conception whereby the Orthodox Christianity is disclosed as an adequate ontological context substantiating both of them. Regarding that, the iconography is considered to be much more than an art being also a science as well as the complex systems also imply an artistic performance as observed in artworks of Jackson Pollock that are construed in terms of the complex systems physics and fractal geometry.

---

[3] The God-man`s personality is revealed by the nature being concurrently human and divine. Stating its complexity the resolution of the Fourth ecumenical council in Chalcedon confesses Jesus Christ '…to be cognized in Two Natures being inconfusable, immutable, indivisible, inseparable; the difference of the Natures being by no mean infringed because of the Union but rather the singularity of each Nature being preserved and concurred to be a Person and a Personality…' (Lossky, 109)

In XX century characterized by radical advances in art, Pollock`s work is considered to be a crucial development (Taylor, 2003, p.118). Described as *all over* style his drip paintings eliminated anything that previously might have been recognized as a composition which made them frequently linked to jazz music. Pollock`s interpretation of his artistic style was about an immediate expression of the nature, concluding that the painter of nowadays cannot express the old form of the Renaissance or any other past culture (Stuckey, 2012). Unlike the modern spatiality rendered by timeless forms the patterns of his artworks are identified as fractal ones and the discovery has since been labelled as *fractal expressionism* (Taylor, 1999; Taylor, 2000).

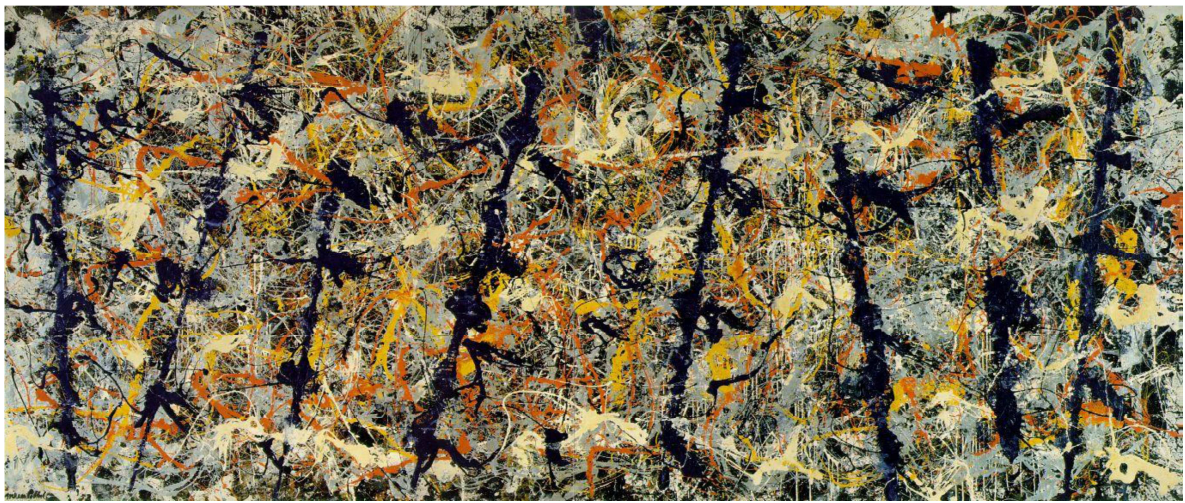

Figure 5. Pollock`s painting *Blue Poles: Number 11, 1952*. The fractal expression of the surface has been linked to a battlefield (Taylor, 2003, p. 118)

The term *fractal* is coined by Benoît Mandelbrot (1967) from a Latin adjective meaning fractured but also irregular in a sense of fragmentation. Despite being like that, the examination of their structure reveals a subtle form of recurrent order. Mandelbrot has demonstrated the scaling property of fractal patterns termed self-similarity that regularly occurs in nature through the same statistical description of a pattern at different resolution scales. It is generally conceived as a generative property related to the growth of an organism in each particular case defining the specific fractal geometry of a matter discussed. Though concerning many biological structures as well as diverse social and cosmic ones, it is manifested in the purest form by the tree rings structure (Fig.6) indicating the intrinsic time of the geometry realized through scaling (Taylor, 1988). Implying 'a time inherent in the texture itself' (Souriau, 1949, p.296-297) fractals are issued as statically impossible features representing an organic access to geometry. Therefore, it is not surprising that Pollock`s artworks were also frequently described as organic ones (Taylor, 2003, p.118) alluding they are patterned by natural phenomena.

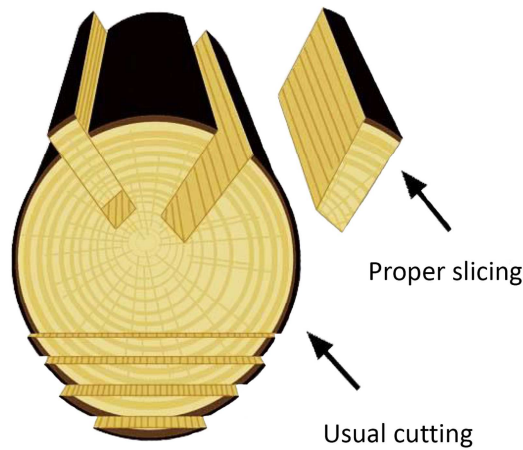

Figure 6. The structure of tree rings representing the intrinsic time through its scaling property. By proper slicing one designates consistence with the tree geometry

The organic process is also evident in the generation of Pollock`s paintings. Because of his unpremeditated style, Pollock was often regarded as an action artist as said by Harold Rosenberg (1952) that the action art had 'broken down every distinction between art and life and that the painting is not art; it is an Is. It is not a picture of a thing: it is the thing itself. It does not reproduce nature, it is Nature'. The idea of painting as an ongoing process clearly reflects the natural process of pattern generation. The so called continuous dynamic process whose pursuer was Pollock, actually concerns self-organization implying an intrinsic time of the art creating inherent to a painting itself. In that regard one conceives his repulsion to sign the paintings when they are completed (Friedman, 1972, p. 185). The individualistic uptake of originality becomes pointless since the artwork is generated through self-organization that is an everlasting process. According to that, the fractal geometry of his artworks is an expression of their originality and is used as such for detecting authenticity of his paintings (Taylor, 2003, p. 141).

The study of human responses to fractal images and the characterization of their appeal is a novel field of research for perceptual psychologists that offers a huge potential (Aks 1996; Briggs, 1992; Geike 1997; Mandelbrot, 1982; Pickover 1995; Rodman, 1957; Rogowitz 1990; Taylor, 1998; Taylor 2003, p. 142). However, one remarks the traditional context of such aesthetics paradigmatically constituted by the Orthodox iconography as have already been mentioned in the paper. In that respect should be figured Pollock`s aesthetical note**:** 'Painting is self-discovery. Every good painter paints what he is,' (O`Conner, 226) concerning a substantial moment of an original art. His conclusion that 'I am nature,' (O`Conner, 226) represents the same aesthetical criterion on the complexity of art and science revealed by a human personality.[4] Out of the ontological context of such aesthetics, covering the art history domain as well as psychology and art theory, a discussion of Pollock`s art should have certainly been incomplete (Taylor, 2003, p.142). Art theorists principally recognize his patterns in terms of a revolutionary approach to aesthetics (Landau, 1989; Varnedoe, 1998).

But the fractal patterns had already been recognized in the iconography (Milovanović, 2013a) almost a half century before the conception was developed and published in

---

[4] Due to an overcome of causal necessity each human personality is revealing the complexity of the nature corresponded to the aesthetical criterion that consolidates both art and science.

Mandelbrot`s (1982) famous book. The geometry was designated by Miloš Radojčić (1940) in terms of a transfiguration power primarily discernible through the form of terrestrial relief,[5] which concerns a substantial conjunction occurring between all creatures and things. In that manner indicating a material personality concurrent to divine one the iconography represents the very core of the Orthodox tradition (Losky, 2003, p.144). As becoming substantiated (in Greek ἐνυπὸστατον[6]) the Nature is figured through a unique aesthetical criterion of both art and science corresponded to communion in immaterial Tabor light (Lossky, 2003, p. 160).

**Conclusion**

In the paper, aesthetics is considered to be an ontological context whose applicability is not restricted to the fine arts domain only. Based upon the originality issue, it impersonates a truth criterion that is an expression of a substantial moment figured through the overcome of causal necessity. The intrinsic time occurring in that respect as the axis of a system self-organization represents the basic concept of such aesthetics. Furthermore, it is argued that the modern science deprived of the aesthetical sensation appeared exactly through elimination of time i.e. its reducing to a linear parameter. Such an uptake of the existence that is characteristic of the modern age has actually denied the traditional personality in the manner of the aesthetical criterion, whose recovery is considered to be a defining feature of the postmodern era. Discussing Kuhn`s (1962) argument on the arbitrariness of aesthetical criteria in science Tibor Machan (1977) wonders what explains that the modern science invokes such a restricted repertoire in their consideration of aesthetical qualities implying simplicity, symmetry and elegance and not qualities that appear far more pervasive in contemporary art. The response regards gnoseological dualism concerning modern humanities due to which both art and science have mislaid a truth criterion in terms of the aesthetical sensation. That is what was appointed by a surrealist artist Hans Arp to be a primordial reality – a mystical urreality that should be re-established by an original artwork.[7]

The exposed criterion impersonates an applied aesthetics related to the art as well as to the physics, biology, cosmology and other developments considered in the complex systems terms. Moreover, it is not about some particular fields of the modern existence but a substantial approach to the existence that is characteristic of the postmodern era overcoming by its transcendence recently present dichotomy between science and religious tradition. Thus, one exactly states the postmodern aesthetics implying by that a criterion that consolidates both art and science through substantiating the nature figured in the light of the Orthodox tradition. Integrating subjective and objective reality by the personality existence, such aesthetics appears as the truth criterion of a substantial gnoseology. The huge potential of its further elaboration concerning ethics and epistemology as well as ecology, culture, education, psychology and other humanities will be discussed elsewhere (Medić-Simić, in preparation).

---

[5] Discussing the fractal geometry Mandelbrot (1967) also starts consideration with elementary geographic instances related to terrestrial relief such as the Britain coastline, the left bank of the Vistula, frontiers of certain countries etc. In that manner should be realized the fractality issue as reassessing the fundamentals of geometry in its primordial significance – a measurement of the earth. His statement that the required self-similarity of the relief cannon yet be fully explained (Mandelbrot, 1975, p.3825) is still relevant.

[6] The term was used by Leontius Byzantinus (Migne, 1865, col. 1277) in order to designate the nature implied by a personality (Lossky, 2003, p. 96).

[7] Arp alludes to time in mythology being integral, original and harmonious. Such aesthetics is a substantial moment of real existence; it is neither fictitious nor arbitrary. Mystical reality having the property of primordial beauty is a prime being that is obscured by an individual subjectivity. The medieval art as a representation of sacral reality was deprived of individually creating arbitrariness (Grasi 1974, p. 90-93).

## Acknowledgments

The authors acknowledge support by the Ministry of Education, Science and Technological Development of the Republic of Serbia through the projects OI 174014 and III 44006, also by the Joint Japan-Serbia Centre for the Promotion of Science and Technology of the University in Belgrade and the ITO Foundation for International Education Exchange.